\documentclass[preprint,12pt]{elsarticle}

\usepackage{graphicx}
\usepackage{color}
\usepackage{multirow}
\usepackage{algorithm}
\usepackage{algorithmic}
\usepackage{hyperref}
\hypersetup{
	colorlinks=true,
	linkcolor=blue
}
\usepackage{adjustbox}

\begin{document}


\begin{frontmatter}
	
	\title{Unmasking unlearnable models: a classification challenge for biomedical images without visible cues}
	\author {Shivam Kumar}
	\author{Samrat Chatterjee\corref{cor1}}
	\cortext[cor1]{Corresponding author: samrat.chatterjee@thsti.res.in}
	\address{Complex Analysis Group, \\ Translational Health Science and Technology Institute, \\ NCR Biotech Science Cluster, \\ Faridabad-121001, India}

\begin{abstract}
Predicting traits from images lacking visual cues is challenging, as algorithms are designed to capture visually correlated ground truth. This problem is critical in biomedical sciences, and their solution can improve the efficacy of non-invasive methods. For example, a recent challenge of predicting MGMT methylation status from MRI images is critical for treatment decisions of glioma patients. Using less robust models poses a significant risk in these critical scenarios and underscores the urgency of addressing this issue. Despite numerous efforts, contemporary models exhibit suboptimal performance, and underlying reasons for this limitation remain elusive. In this study, we demystify the complexity of MGMT status prediction through a comprehensive exploration by performing benchmarks of existing models adjoining transfer learning. Their architectures were further dissected by observing gradient flow across layers. Additionally, a feature selection strategy was applied to improve model interpretability. Our finding highlighted that current models are unlearnable and may require new architectures to explore applications in the real world. We believe our study will draw immediate attention and catalyse advancements in predictive modelling with non-visible cues. 
Our source code is available at \url{https://github.com/samrat-lab/Image-classification-3d}.
\end{abstract}

\end{frontmatter}
\section{Introduction}
\label{sec:intro}

Radiogenomics is a rapidly evolving field, and it is defined as the association of imaging phenotype with genomic characteristics \cite{pinker2018background}. These could be the collective expression pattern of genes or any individual mutations. One of the most useful applications is precision medicine, which has been proven saviour in recent years for complex diseases like cancer \cite{dohner2021towards}. Conventional practices bottleneck this technique's growth, including identifying genetic markers by invasive processes, including tissue extraction, doing relevant assays, and waiting for the result. These limitations are being overcome by radiomics, which includes binary classification of certain genotypes like IDH and 1p arm co-deletion using MRI/CT images \cite{ding2019prediction,yan2022predicting}. 
Advanced imaging algorithms, such as deep convolutional networks, have demonstrated promising outcomes in this context. Such predictive models hold significant potential in biology, particularly in advancing precision medicine towards non-invasive methodologies \cite{johnson2021precision,meng2019preoperative,zhao2022development}. However, while some results have shown initial promise of above 90\% \cite{chakrabarty2023mri,li2018genotype}, reproducibility across independent cohorts has posed a challenge by lowering accuracy to below 80 \% \cite{ccelik2021magnetic,pasquini2021deep,guo2023noninvasive,zhou2021deep,kandemirli2021machine,pan2019machine}. 
Experts annotate these labels visually in many computer vision tasks, and the machine replicates this behaviour. However, the input data and corresponding labels are gathered from separate sources in biomedical sciences. While the input data is generated through regular imaging, the corresponding labels often stem from invasive procedures because the visual characteristics necessary for annotation are often unknown or not readily discernible. So, predicting subtle signatures that are not prominently visible in images poses a significant challenge.

To understand the problem in detail, we chose to predict MGMT methylation status from 3D MRI images of glioma patients. MGMT, a crucial marker, is essential for choosing chemotherapy to increase patient survivability \cite{rivera2010mgmt}. Glioma images show subtle visible characteristics like diffusion, which could be linked to MGMT status and help clinicians to know its status through non-invasive methods. 

Several studies have been reported in the literature for predicting MGMT status from MRI images; most sources came from imaging archives of the cancer genome atlas, and some of the studies had their in-house dataset \cite{capuozzo2022multimodal,lost2023systematic,kihira2022u}. Thus far, these reported studies have predictive performance with a low accuracy. Until recently, RNSA-BRATS launched one of the most extensive and standardized datasets and a public challenge of radiogenic prediction in 2021. Several solutions have been provided, and the leaderboard result reported a cumulative AUROC of 0.6, which included a ResNet model with ten layers. Following the closure of this competition and brats-2021 being one of the most extensive publicly available datasets for MGMT, a cumulative effort by the community has been made to improve the performance \cite{das2022optimizing,qureshi2023radiogenomic,mohamed2024brain,qu2022attentive,kollias2023btdnet,hu2024mgmt,han2024synthesis,palsson2021prediction}.
Despite numerous efforts, contemporary models exhibit suboptimal performance, and underlying reasons for this limitation still need to be discovered. Many studies do not support the association between MGMT status and MRI images \cite{saeed2022possible,kim2022validation,saeed2023mgmt,sakly2023brain,robinet2023mri}.They computationally predicted no correlation between MGMT status and MRI images, which is a setback for radiogenic study for MGMT. On the other hand, some studies on their internal dataset have reported some extent of prediction \cite{capuozzo2022multimodal,lost2023systematic,kihira2022u}. Reaching such a conclusion without an in-depth computational investigation would be too soon. The main reason for such inconclusive outcomes is the lack of predictive models on the images without visual cues. This understanding motivated us to dig deeper towards the computational task of predicting phenotype from such images. We aim to study this problem through a process-driven approach to identify the necessity of indigenous tools in this domain. In this study, we demystify the complexity of MGMT status prediction through a comprehensive exploration by performing benchmarks of existing models adjoining transfer learning. We also studied gradient flow across layers and applied a feature selection strategy. The study ends with the possible reason behind the persistent saturation and discusses possible solutions to such problems.

\section{Methods}

\subsection{Dataset}

The dataset used in this study is from the Brain Tumor Radiogenomic Classification challenge (Brats-2021) \cite{baid2021rsna}, consisting of MRI scans for 585 glioblastoma patients. The patients belong to unmethylated MGMT and methylated MGMT, containing 278 and 307 samples, respectively. Scanned images are in four different modalities: T1-weighted pre-contrast (T1), T1-weighted post-contrast (T1CE), T2-weighted (T2) and T2 Fluid attenuated Inversion Recovery (FLAIR). This study will use all imaging modalities for most of the exercise until otherwise mentioned. The brain images provided are already preprocessed by resampling and skull stripping.

\subsection{Convolutional neural network, training and evaluation strategy}

The standard Convolutional neural network (CNN) architecture like ResNet \cite{he2016deep}, DenseNet \cite{huang2017densely} and EfficientNet \cite{tan2019efficientnet} has been opted for evaluating the predictive models. The choice of this baseline is supported by their success in many biomedical classifications and the robustness of their performance. The optimizers are Adam and RMSprop(RMS), and the loss functions for classification tasks are Hinge and BCElogitloss. A batch size of 4 was utilized \cite{kollias2023btdnet}, alongside a learning rate of 0.0001, with epochs ranging from 25 to 100 in certain scenarios. Image dimensions were set to 256$\times$256 \cite{mohamed2024brain}. Assessment of the models' predictive efficacy was done using accuracy, AUROC, sensitivity, and specificity.

\subsection{Transfer learning}

Transfer learning is an approach where we use weight from previously published models and extend the training process with our data. Recently, this approach has been proven effective in the case of less data and dense architecture training \cite{weiss2016survey}. In our study, we used ImageNet and MedNet weight for the model weight initialization \cite{deng2009imagenet,chen2019med3d}. Later, we used two transfer approaches: the first using previously trained models as feature extractors and training the rest of the fully connected layers. In the second approach, we used fine-tuning, i.e., we trained the complete architecture with initial weights from MedNet.

\subsection{Explainable approach to model improvement}

To understand possible ways to enhance the predictive capability of current models, we adopted an explainable approach consisting of several key steps. First, we use pyradiomics \cite{van2017computational} to extract first-order statistics (9 features), Shape-based (2D) features (10 features), Gray Level Co-occurrence Matrix features (24 features), Gray Level Run Length Matrix( 16 features), Gray Level Size Zone Matrix feature (16 features), Neighbouring Gray Tone Difference Matrix features (5 features), and Gray Level Dependence Matrix (14 features) from the MRI images. Subsequently, we conducted the Mann-Whitney U test to assess whether the distributions of these parameters varied significantly. 
Later, features with p-value $\leq$ 0.05 were deemed statistically significant and retained for further analysis. To identify the most informative features, we employed recursive feature elimination using a random forest algorithm (Algorithm \ref{algo1}). The top-ranked features were then selected to construct a predictive model, whose performance was evaluated against an initial set of significant features. Additionally, we employed hierarchical clustering to group features based on their similarities. This clustering analysis provided insights into the importance of different feature groups and guided the development of novel architecture.
\begin{algorithm}[h]
	\caption{Recursive elimination of features with cross-validation}
	\label{algo1}
	\begin{algorithmic}[1]
		\STATE \textbf{Input:} Samples with $N$ features
		\STATE \textbf{Output:} A set $F_{master}$ having number of feature used to obtain a specific accuracy 
		\STATE $S \gets Features\_Initial$
		\STATE Build RF model on $S$ and rank features using model coefficients
		\STATE $S_{rank} \gets$ sorted feature from high to low based on gini index
		\FOR{$i$ = Num\_Features  \TO $1$}
		\STATE $f_{least} \gets S_{rank}[i]$
		\STATE $S_{acc} \gets $ accuracy using cross-validation on $S - \{f_{least}\}$
		
		\STATE $S_{new} \gets S - \{f_{least}\}$ 
		\STATE Store $(S_{new},S_{acc})$ to $F_{master}$
		\ENDFOR
		
	\end{algorithmic}
\end{algorithm}

\section{Experiments}

\subsection{Predictive performance of various CNN models}

Building upon previous investigations of MGMT and reproducing previous studies \cite{saeed2023mgmt} is a crucial step for fair investigation of models. The performance of the three CNN models— DenseNet264, ResNet101, and EfficientNet—was evaluated across the entire dataset encompassing all four MRI image types (Table \ref{table_feat}). The accuracy on all the image types ranges between 49 \%-63\%. It was hard to infer that a specific image modality gave the highest performance because, for fold 1, FLAIR had the highest accuracy. However, for fold 3, T2 showed maximum performance using DenseNet and similar behaviour was observed for others on specific folds. Despite employing complex architectures and extensive training, the models exhibited sub-optimal predictive accuracy, underscoring the challenges inherent in this predictive task. Due to inconsistent predictive accuracy, these metrics are not concrete enough to make plausible decisions about the model selection for downstream analysis. So, we chose ResNet as it is small in parameter size among these models and does not require heavy computational resources to weigh out different parameter possibilities. Further, we have chosen the T2 image type, as it is better at capturing the tumour and its character \cite{bladowska2011t2}.
\begin{table}[h]
	\begin{tabular}{|l|llll|l|llll|}
		\hline
		&
		\multicolumn{4}{l|}{DenseNet} &
		&
		\multicolumn{4}{l|}{ResNeT} 
	\\ \hline
		&
		\multicolumn{1}{l|}{FLAIR} &
		\multicolumn{1}{l|}{T1w} &
		\multicolumn{1}{l|}{T2w} &
		T1wCE &
		\multirow{6}{*}{} &
		\multicolumn{1}{l|}{FLAIR} &
		\multicolumn{1}{l|}{T1w} &
		\multicolumn{1}{l|}{T2w} &
		T1wCE \\
		\cline{1-5} \cline{7-10} 
		Fold1 &
		\multicolumn{1}{l|}{62.4} &
		\multicolumn{1}{l|}{54.7} &
		\multicolumn{1}{l|}{59} &
		51.3 &
		&
		\multicolumn{1}{l|}{56.4} &
		\multicolumn{1}{l|}{58.1} &
		\multicolumn{1}{l|}{59} &
		55.6 
		\\ \cline{1-5} \cline{7-10} 
		Fold2 &
		\multicolumn{1}{l|}{61.5} &
		\multicolumn{1}{l|}{60.7} &
		\multicolumn{1}{l|}{58.1} &
		54.7 &
		&
		\multicolumn{1}{l|}{58.1} &
		\multicolumn{1}{l|}{56.4} &
		\multicolumn{1}{l|}{56.4} &
		53
	\\ 
	\cline{1-5} \cline{7-10} 
		Fold3 &
		\multicolumn{1}{l|}{52.1} &
		\multicolumn{1}{l|}{49.6} &
		\multicolumn{1}{l|}{58.1} &
		54.7 &
		&
		\multicolumn{1}{l|}{59.8} &
		\multicolumn{1}{l|}{53.8} &
		\multicolumn{1}{l|}{54.7} &
		55.6 
	\\ \cline{1-5} \cline{7-10} 
		Fold4 &
		\multicolumn{1}{l|}{57.3} &
		\multicolumn{1}{l|}{57.3} &
		\multicolumn{1}{l|}{59.8} &
		54.7 &
		&
		\multicolumn{1}{l|}{50.4} &
		\multicolumn{1}{l|}{59.8} &
		\multicolumn{1}{l|}{63.2} &
		51.3 
	\\ \cline{1-5} \cline{7-10} 
		Fold5 &
		\multicolumn{1}{l|}{53.8} &
		\multicolumn{1}{l|}{53} &
		\multicolumn{1}{l|}{54.7} &
		52.1 &
		&
		\multicolumn{1}{l|}{59.8} &
		\multicolumn{1}{l|}{50.4} &
		\multicolumn{1}{l|}{54.7} &
		52.1 
	\\ \hline
	\end{tabular}
\begin{tabular}{|l|llll|}
	\hline
	&
	\multicolumn{4}{l|}{EfficientNET} \\ \hline
	\multirow{6}{*}{} &
	\multicolumn{1}{l|}{FLAIR} &
	\multicolumn{1}{l|}{T1w} &
	\multicolumn{1}{l|}{T2w} &
	T1wCE \\ \cline{1-5} 
	Fold1 &
	\multicolumn{1}{l|}{53.8} &
	\multicolumn{1}{l|}{56.4} &
	\multicolumn{1}{l|}{52.1} &
	60.7 \\ \cline{1-5} 
	Fold2 &
	\multicolumn{1}{l|}{56.4} &
	\multicolumn{1}{l|}{60.7} &
	\multicolumn{1}{l|}{50.4} &
	55.6 \\ \cline{1-5} 
	Fold3 &
	\multicolumn{1}{l|}{58.1} &
	\multicolumn{1}{l|}{50.4} &
	\multicolumn{1}{l|}{54.7} &
	55.6 \\ \cline{1-5} 
	Fold4 &
	\multicolumn{1}{l|}{55.6} &
	\multicolumn{1}{l|}{56.4} &
	\multicolumn{1}{l|}{57.3} &
	56.4 \\ \cline{1-5} 
	Fold5 &
	\multicolumn{1}{l|}{58.1} &
	\multicolumn{1}{l|}{54.7} &
	\multicolumn{1}{l|}{53.8} &
	52.1 \\ \hline
\end{tabular}
	\vspace{0.2cm}
	\caption{Accuracy for five fold testing using resnet, densenet and effiecientnet for all four image modalities (FLAIR, T1w, T2w, T1wCE).}
	\label{table_feat}
\end{table}

\begin{figure}[h]
	\includegraphics[width=\textwidth]{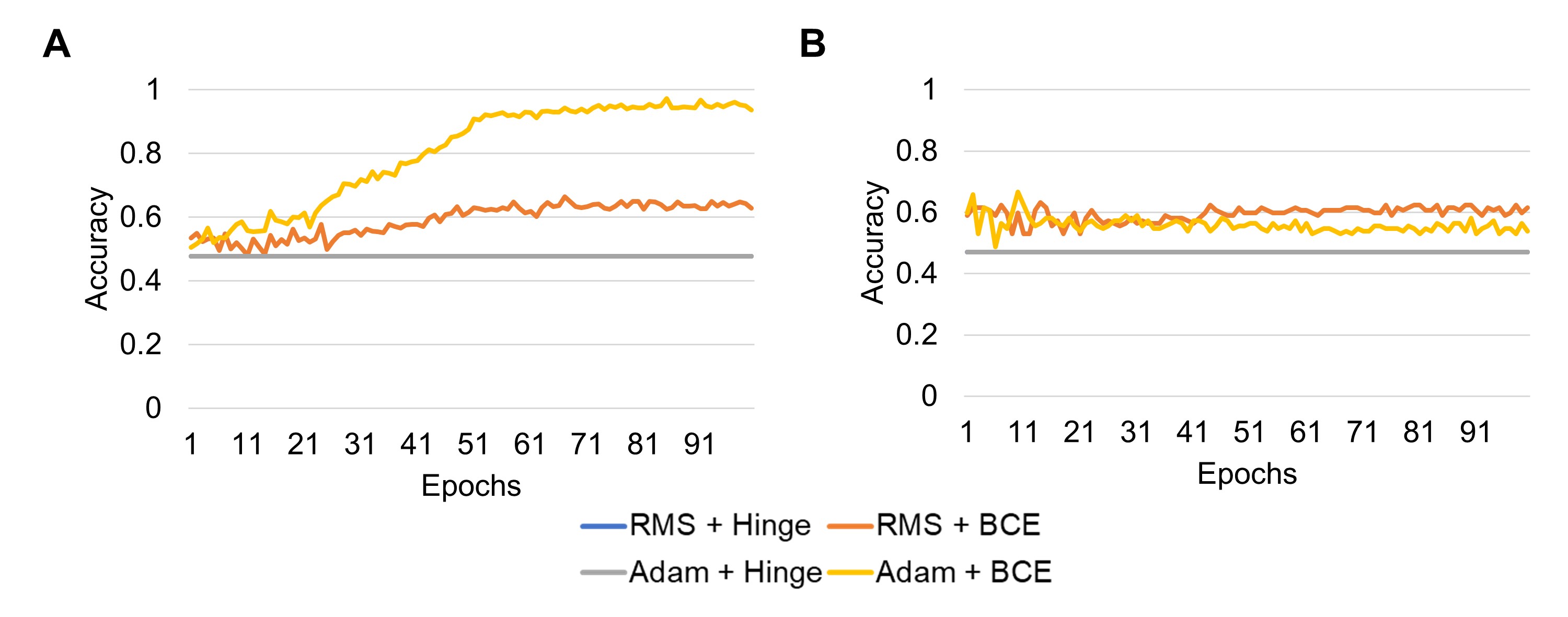}
	\caption{\textbf{Comparison of hyper parameters} (A) Testing different combinations of loss function and error metric on training. (B) Testing different combinations of loss function and error metric on test dataset. In both the cases the accuracy for Hinge with RMS and Adam has been overlapped. }
	\label{fig1}
\end{figure}

Additionally, to get the best performance of ResNet, we tested the combination of error metrics and optimizers to discern their impact on predictive performance (Figure \ref{fig1}). The results show that Hinge with RMS and Adam has a flat and overlapping line of accuracy throughout the epoch, suggesting random predictive behaviour. BCE with RMSProp showed slight improvement with saturation at 60\% accuracy after 50 epochs. BCE  showed steep monotonicity till 50 epochs and reached 95\% till ten epochs when complimented with Adam. However, the testing dataset did not show this behaviour; the maximal performance was 60\% in this case.

The performance of the task remained maximum at 60\% in a single split, despite implementing complex models \cite{he2016deep,huang2017densely,tan2019efficientnet}. Adjusting parameters such as loss functions and optimizers also failed to yield significant enhancements. Given the complexity of these models, training them on small datasets of this magnitude may hinder their ability to capture meaningful patterns effectively. So, researchers have turned to a transfer learning approach to improve poor-performing models \cite{vasquez2021transfer}. 

\subsection{Transfer learning approach for improving model accuracy}

\begin{figure}[h]
	\includegraphics[width=\textwidth]{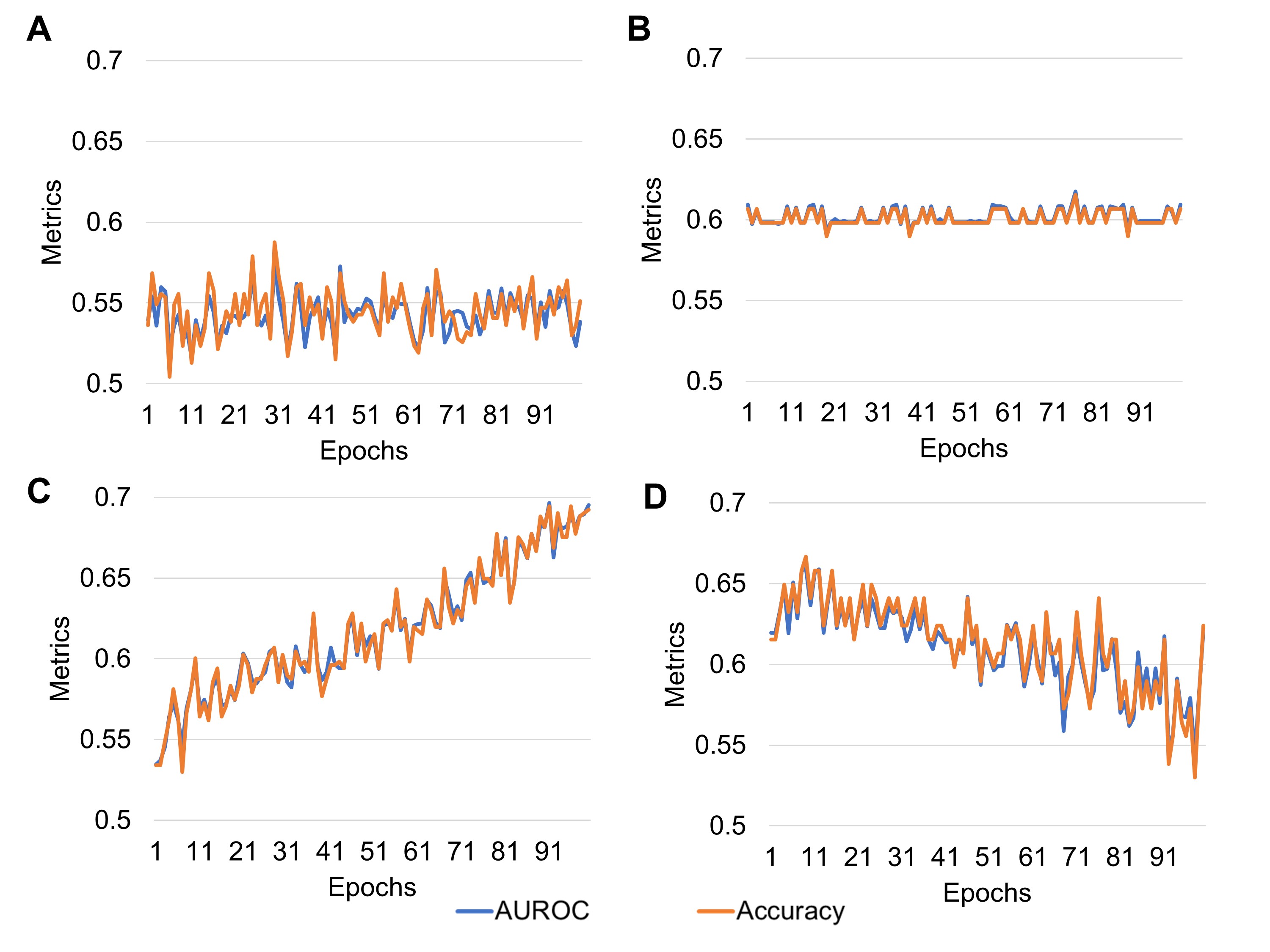}
	\caption{\textbf{The various transfer learning approach using MedNet weights} (A-B) The MedNet weight used as feature extractor and fully connected layer are trained using current data in training and test split. (C-D) The MedNet weight was fine-tuned using the current dataset and performance evaluated on training and test data.}
	\label{fig2}
\end{figure}

The two kinds of transfer learning approaches have been implemented. The first one is the interdomain, where we used ImageNet weight, and the intradomain MedNet weights were used. Initially, the ResNet10 model was utilized as a feature extractor to obtain meaningful representations from MRI images for predicting MGMT status (Figure \ref{fig2} A-B). In this process, the model's weights were fixed, and only the fully connected layers were trained. The training or test datasets did not significantly improve predictive performance despite training the model over multiple epochs. This outcome suggests that the features learned by the model could not correlate with output labels effectively. The finding highlights that the cross-task feature transfer is ineffective when the nature of the task is challenging, like finding subtle patterns.

Next, we fine-tuned the ResNet architecture with pre-initialized MedNet weights to improve the model's performance potentially. The complete network was trained using the Brats dataset (Figure \ref{fig2} C-D) for fine-tuning. The model resulted in noticeable enhancements (up to 70\%) in predictive accuracy on the training data, indicating that it adapted its features more effectively to the task. However, it is worth noting that the test dataset's performance remained unchanged over the first 20 epochs and later dipped below 60\%. It is unclear whether the model captured specific patterns related to MGMT status but over-fitted so well that it could not generalize its performance on test data. There is also the possibility of memorizing information without capturing any pattern.
Further investigation into model behaviour using parameters like sensitivity and specificity becomes crucial in the learning curve in such scenarios. Additionally, analyzing the gradient flow is pertinent, as it helps determine if the model is learning meaningful patterns. If the weights are consistently updated across epochs, eventually reaching a plateau, the model may have learned all it can from the data; then, this becomes a generalization problem; otherwise, it is a pattern-recognizing problem.

\subsection{Model investigation for saturated accuracy}

\subsubsection{Analyzing learning curve with sensitivity and specificity}

\begin{figure}[h]
	\includegraphics[width=\textwidth]{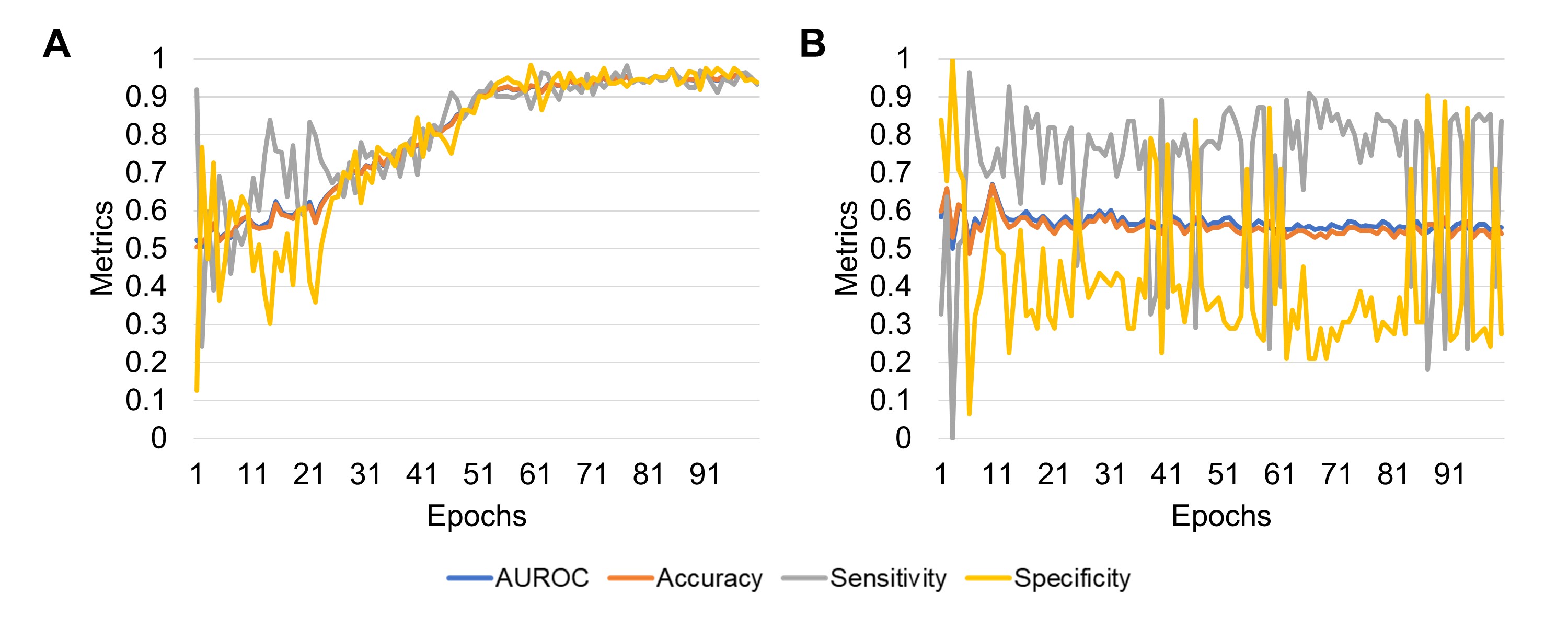}
	\caption{\textbf{The model investigation using learning curve with sensitivity and specificity.} (A-B) The fluctuation of sensitivity and specificity in training and test dataset.}
	\label{fig3_1}
\end{figure}

A learning curve represents the relationship between a model's performance (often measured by accuracy) and the number of training epochs. As the model undergoes more training epochs, its accuracy on the training data tends to increase, which we have observed here (Figure \ref{fig3_1}).

To understand each category's learning behaviour, we added two more metrics: sensitivity and specificity. The result shows that the sensitivity and specificity were showing complementary and fluctuating behaviour in the early epochs of the training dataset (Figure \ref{fig3_1} A) and entire epochs of the test dataset (Figure \ref{fig3_1} B).
This erratic behaviour explains the saturation in the accuracy, where sudden spikes or drops and plateaus were observed, corresponding to instances where the model predominantly predicted one class while ignoring the other. Consequently, this insight suggests that the observed lower accuracy of ~60\% is not due to the learning behaviour but random favour causing either lower sensitivity or specificity.

\subsubsection{Inspecting the gradient flow in training epochs}

\begin{figure}[h]
	\includegraphics[width=\textwidth]{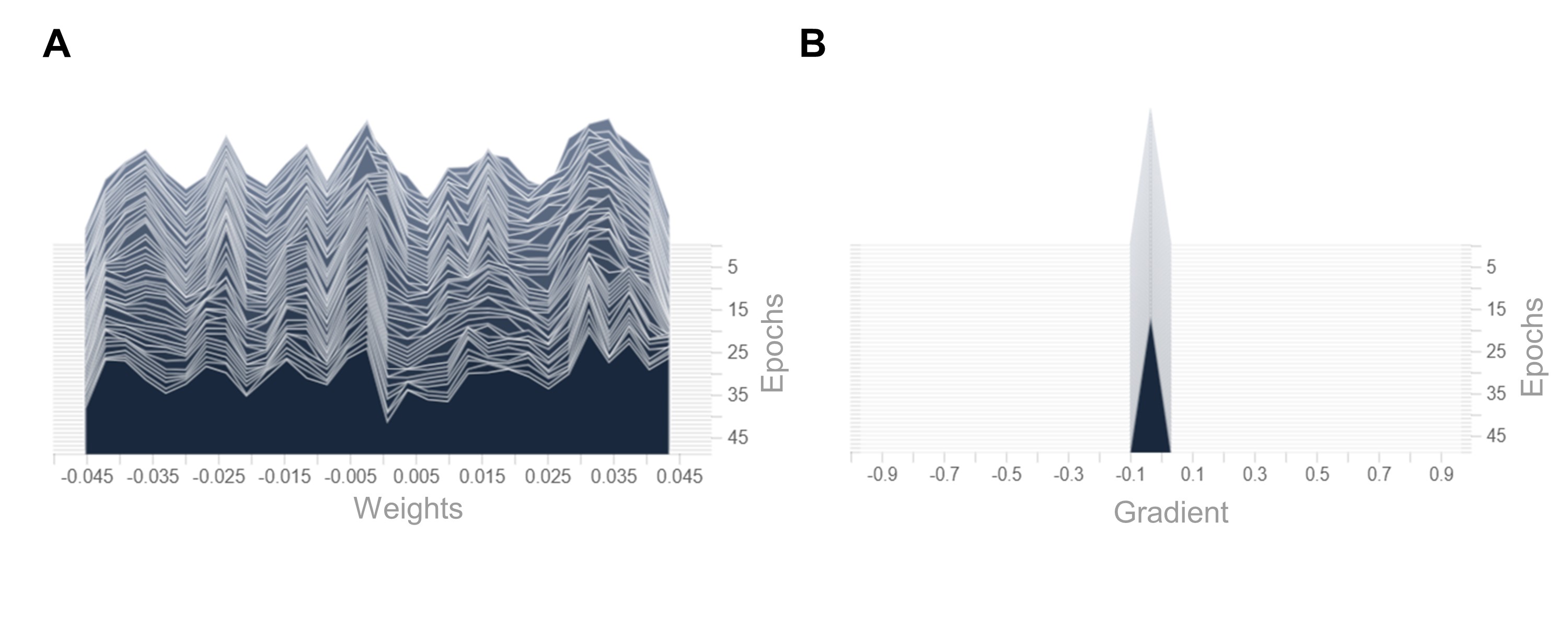}
	\caption{\textbf{The model investigation using gradient flow.} (A) The model weight in the primary layer of ResNet.  The figure shows the histogram for different epochs with learned weights. (B) The gradient updates for subsequent epochs of training. The figure shows the histogram for different epochs with calculated gradient of weights.}
	\label{fig3_2}
\end{figure}

Here, we examined the gradient flow within the model to investigate the observed random favour. We have shown a histogram of the cumulative model weight of the initial layer (Figure \ref{fig3_2} A) and its corresponding gradient (Figure \ref{fig3_2}) for every epoch. The result shows that the model weights remained unchanged despite training over multiple epochs. This lack of weight updates suggested that the model was not effectively learning from the training data, leading to suboptimal performance and erratic predictions. We also observed that certain model layers were not receiving new gradients during backpropagation (Figure \ref{fig3_2} B), resulting in minimal weight updates. This phenomenon was mainly observed in all the layers but is prominent in the initial layer, which captures major imaging features. The above investigation, thus, shows that the models need to learn the effective behaviour required for predictions. 

In the literature, it was observed that sometimes increasing model complexity might improve the performance of the prediction algorithm \cite{yamanaka2017fast} and so it leads to the following analysis.

\subsection{Saturated performance and model complexity}

To obtain the relationship between complexity and model performance, we increase the complexity of ResNet by adding convolution layers in three sets, namely 10, 34, and 50 \cite{wang2020multi}. The result shows no clear dependence between performance in terms of accuracy and AUROC (See Table \ref{table2}). Both low and high-complexity models gave accuracy between 52-60\%, showing that the model accuracy is independent of its complexity.

\begin{table}[h]
	\centering
	\begin{minipage}{0.9\textwidth}
		\centering
	\begin{tabular}{|l|lll|l|lll|}
		\hline
		& \multicolumn{3}{l|}{Accuracy}                                &  & \multicolumn{3}{l|}{AUROC}                                      \\ \hline
		& \multicolumn{1}{l|}{ResNet10} & \multicolumn{1}{l|}{ResNet34} & ResNet50 & \multirow{6}{*}{} & \multicolumn{1}{l|}{ResNet10} & \multicolumn{1}{l|}{ResNet34} & ResNet50 \\ \cline{1-4} \cline{6-8} 
		Fold1 & \multicolumn{1}{l|}{55.5} & \multicolumn{1}{l|}{57.2} & 57.2 &  & \multicolumn{1}{l|}{0.513} & \multicolumn{1}{l|}{0.573} & 0.526 \\ \cline{1-4} \cline{6-8} 
		Fold2 & \multicolumn{1}{l|}{65.8} & \multicolumn{1}{l|}{59.8} & 60.6 &  & \multicolumn{1}{l|}{0.626} & \multicolumn{1}{l|}{0.617} & 0.618 \\ \cline{1-4} \cline{6-8} 
		Fold3 & \multicolumn{1}{l|}{53.8} & \multicolumn{1}{l|}{55.5} & 55.5 &  & \multicolumn{1}{l|}{0.538} & \multicolumn{1}{l|}{0.534} & 0.533 \\ \cline{1-4} \cline{6-8} 
		Fold4 & \multicolumn{1}{l|}{58.1} & \multicolumn{1}{l|}{56.4} & 53.8 &  & \multicolumn{1}{l|}{0.545} & \multicolumn{1}{l|}{0.482} & 0.462 \\ \cline{1-4} \cline{6-8} 
		Fold5 & \multicolumn{1}{l|}{57.2} & \multicolumn{1}{l|}{58.1} & 52.1 &  & \multicolumn{1}{l|}{0.524} & \multicolumn{1}{l|}{0.555} & 0.437 \\ \hline
	\end{tabular}
	\vspace{0.2cm}
	\caption{Relation between different model complexities and predictive performances in terms of accuracy and AUROC.}
	
	\label{table2}
\end{minipage}
\end{table}


Our understanding so far led us to conclude that current models are unlearnable and fail to incorporate imaging features. Moreover, we may not need a bigger and more complex model. Simple architectures could also work in these tasks, provided they capture the subtle patterns available in the image. As these patterns are not prominent, we can guide the algorithm with specific knowledge, which could be domain-driven to explore possible ways to improve the current regime.

\subsection{Extracting explainable radiomic feature for the solution}

\begin{figure}
	\includegraphics[width=\textwidth]{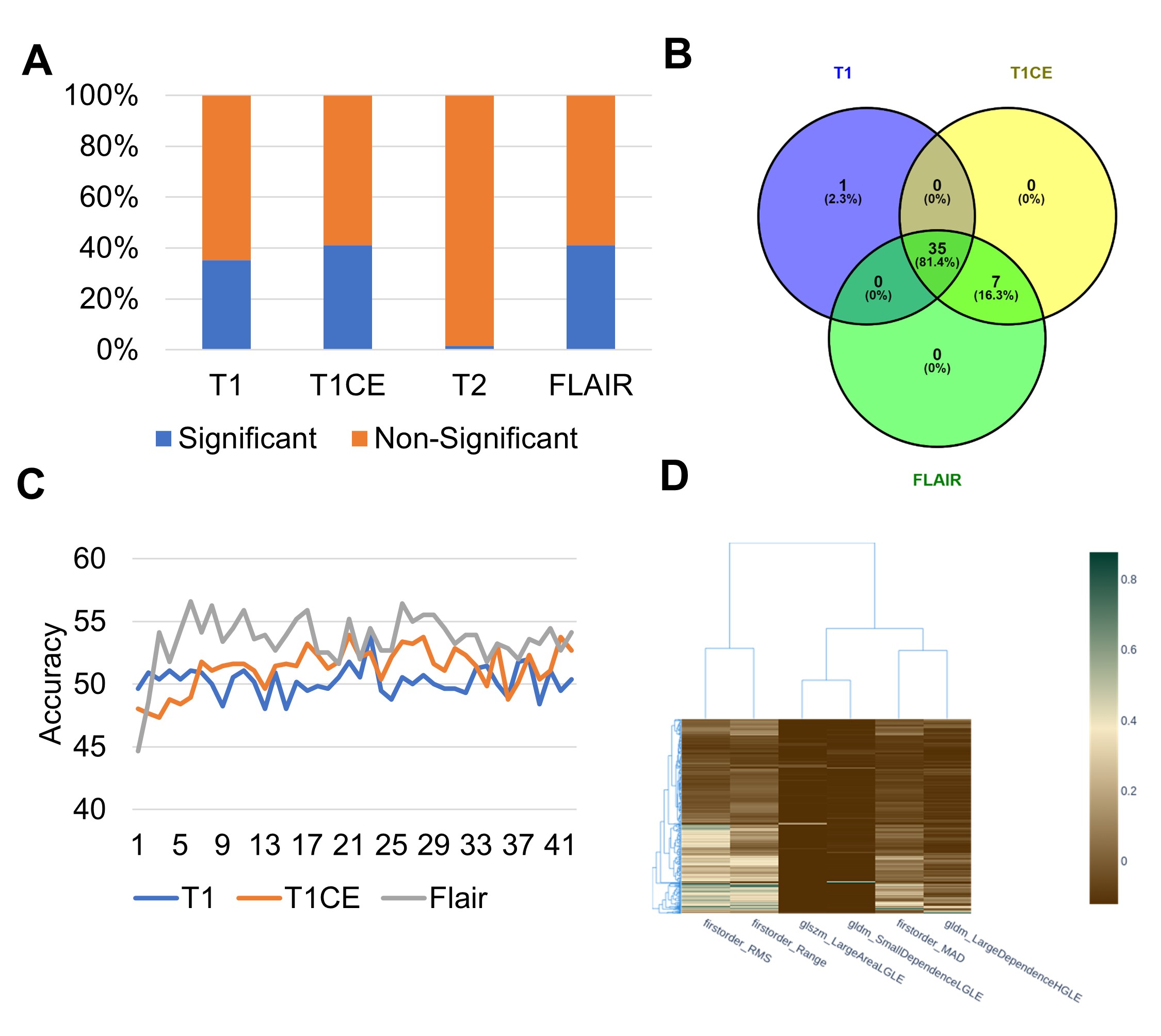}
	\caption{\textbf{Determining important radiomic feature between MGMT methylated and non-methylated category.} (A) Statistically significant features for different image modalities. (B) Common feature among the image modality (T1, T1CE, FLAIR), Here image modality T2 has not been considered as it has only one significant feature, which was not common in any other modality. (C) Feature ranking using recursive feature elimination with random forest. (D) Clustering of similar feature using dendrogram, showing three clusters with two features each. }
	\label{fig5}
\end{figure}

Building upon our exploration of various CNN architectures and training methodologies, it has become evident that the subtle patterns in the images pose a formidable challenge. The generic feature maps generated by CNNs may not fully capture the intricacies present. So, to complement the CNN, we turned to radiomics features, which offer invaluable insights into the underlying characteristics of the images. These features might serve as a window into the black box, allowing us to explore the reasons behind model weaknesses and avenues for improvement. The statistically significant features across modalities were captured in figure \ref{fig5} A. The T2 modality has only one significant feature, while FLAIR reveals 42 significant features (p-val$\leq$0.05) among 104.
Similarly, T1 and T1Ce modalities yield 36 and 42 significant features, respectively. To understand whether the feature depends on image modality, we highlighted significant features among FLAIR, T1, and T1Ce modalities (Figure \ref{fig5} B). FLAIR shares 43 features with T1Ce, 35 with T1, and the latter two share 35 features. Across all modalities, 35 features were common, with a total of 43. We have implemented feature selection using recursive elimination to choose the best among them.
Consequently, we found that each image requires different features to achieve maximum performance. FLAIR, T1, T1CE and T2 selected 6, 23, 8 and 1 features, respectively (Figure \ref{fig5} C). Among them, FLAIR gave the highest performance with six features. So, to assess the relevance of essential features, we have compared them with all 43 features. The prediction model for all and top important features shows that feature selection does improve the model performance  (Table \ref{acc_table}). However, these metrics still need to be higher, suggesting that while these features are essential, they may not suffice for real-world applications alone.

\begin{table}[h]
		\centering
	\begin{minipage}{0.6\textwidth}
		\centering
	\begin{tabular}{|l|l|l|}
		\hline
		& All-Feature & Top-Feature \\ \hline
		Accuracy  & 0.513       & 0.559       \\ \hline
		F1-Score  & 0.533       & 0.582       \\ \hline
		AUROC     & 0.52        & 0.58        \\ \hline
		Precision & 0.537       & 0.582       \\ \hline
		Recall    & 0.535       & 0.588       \\ \hline
	\end{tabular}
\vspace{0.2cm}
\caption{Predictive model performance with all features and top features.}
\label{acc_table}
\end{minipage}
\end{table}

These identified features are important but it is crucial to understand their association. We have observed top features using a dendrogram, showing three clusters with two features each (Figure \ref{fig5} A). So these three groups can be used to derive custom feature maps. Implementing knowledge driven feature maps can enhance the learning of convolution networks, but caution is warranted to mitigate biases and ensure generalization ability across diverse datasets.

\section{Conclusion and Future direction}

The paper aims to dig deeper towards the computational task of building predictive models using images with non-visible cues. Majorly, solutions to such problems are given by techniques not designed for solving such problems due to outcomes-driven emphasis, which requires an in-depth exploration to identify the necessity of indigenous tools in this domain. To understand the problem through a process-driven approach, we chose a case study of predicting MGMT methylation status from MRI images. We showed that the performance of the models remained unchanged despite various adjustments, including alterations to the loss function and the exploration of transfer learning techniques. Further examination of the learning curve revealed that the models exhibited poor learnability, indicating challenges in their training process. We conclude that it is not fair to use techniques designed to capture strong visible patterns to recognize subtle minor variances that are not observed visually. It is well known in the literature that multiple weak learners could aggregate to one strong learner \cite{feng2020machine}. So, we sought help with an explainable approach, which provided insights into potential avenues for enhancing model architectures. Adding some influential radiomic features may enhance the performance. However, it may also create bias and require further exploration. The current study aims to draw attention to the sensitivity of the problem, so instead of providing a concrete solution, we limit this study only to a possible direction towards the solution. In future research, we plan to develop custom feature maps informed by the extracted knowledge of essential features, thereby addressing the underlying identified issues.



\bibliographystyle{unsrt}
\bibliography{Reference}
\end{document}